# Beyond the Screen: Safeguarding Mental Health in the Digital Workplace Through Organizational Commitment and Ethical Environment

**Ali Bai**

Wilson College of Business, University of Northern Iowa, Cedar Falls, Iowa, USA

ali.bai@alumni.uni.edu

**Morteza Vahedian**,

Department of Sociology, College of Humanities, University of Kashan, Kashan, Iran

mvahedian@grad.kashanu.ac.ir

**Abstract**

This research explores the intricate relationship between organizational commitment and nomophobia, illuminating the mediating influence of the ethical environment. Utilizing Meyer and Allen's three-component model, the study finds a significant inverse correlation between organizational commitment and nomophobia, highlighting how strong organizational ties can alleviate the anxiety of digital disconnection. The ethical environment further emerges as a significant mediator, indicating its dual role in promoting ethical behavior and mitigating nomophobia's psychological effects.

The study's theoretical advancement lies in its empirical evidence on the seldom-explored nexus between organizational commitment and technology-induced stress. By integrating organizational ethics and technological impact, the research offers a novel perspective on managing digital dependence in the workplace. From a practical standpoint, this study serves as a catalyst for organizational leaders to reinforce affective and normative commitment, thereby reducing nomophobia. The findings underscore the necessity of ethical leadership and comprehensive ethical policies as foundations for employee well-being in the digital age.

Conclusively, this study delineates the protective role of organizational commitment and the significance of ethical environments, guiding organizations to foster cultures that balance technological efficiency with employee welfare. As a contribution to both academic discourse and practical application, it emphasizes the importance of nurturing a supportive and ethically sound workplace in an era of pervasive digital integration.

# Introduction

In the intricate landscape of organizational behavior, the emergence of 'nomophobia'—the trepidation of being without a mobile phone—reflects a deeper narrative about our technological dependencies and their intersection with workplace dynamics. This contemporary ailment, symptomatic of our digital age, is not merely a personal inconvenience but resonates through the fabric of organizational culture, potentially influencing job satisfaction and employee performance. The phenomenon of nomophobia, particularly prevalent in today's hyper-connected society, demands a comprehensive understanding of the context of organizational structures and the ethical environments they foster.

Organizational commitment, a multifaceted construct that reflects an employee's psychological attachment to their organization, has long been studied for its implications on work behavior and attitudes. The seminal three-component model proposed by Meyer and Allen (1991) provides a robust framework to explore this construct, encompassing affective, continuance, and normative commitment. These facets of commitment offer a lens through which to assess how employees' allegiance to their organization may buffer against or exacerbate the effects of nomophobia.

The ethical environment within organizations, defined by the collective perception of normative conduct, is another cornerstone of this discussion. A sound ethical climate has been identified as a catalyst for positive workplace behavior, counteracting job-related anxieties and bolstering job satisfaction (Schwepker, 2023). Bai et al. (2023) emphasize the dual influence of spiritual intelligence and ethical environments in enhancing women's job satisfaction, suggesting that such environments may serve as a mediating factor in the relationship between organizational commitment and nomophobia.

At the intersection of these variables lies the broader organizational context, where the consequences of nomophobia extend to facets such as productivity, cybersecurity, and work-life balance (Gupta, Agrawal, & Gaur, 2021). The challenge for contemporary organizations lies not only in leveraging technology for enhanced business outcomes but also in safeguarding the psychological well-being of their workforce in an always-connected world.

This research aims to bridge the gap in the literature by exploring the impact of organizational commitment on nomophobia and the mediating effect of the ethical environment. By integrating theoretical perspectives from organizational behavior, information systems, and ethics, the paper seeks to provide a holistic view of how modern technology impacts employee attachment and the ensuing organizational implications.

Recent studies have further nuanced our understanding of nomophobia within the organizational context. Hessari et al. (2022a) explore the role of supportive leadership and co-worker support in mitigating nomophobia, emphasizing the importance of affective commitment and HRM practices in reducing its prevalence. This is complemented by Hessari and Nategh's (2022b) investigation into the dual nature of smartphone addiction, which can both maximize and minimize job performance through the mediating variables of life invasion and techno exhaustion. Moreover, their subsequent work (2022c) delves into the role of co-worker support in tackling technostress and its subsequent effects on employees' need for recovery and work motivation, underscoring the multifaceted impact of technology on the modern workforce.

## Theoretical Development

### Nomophobia

Organizations are indelibly interlinked with technology, such as using workplace Internet connectivity, desktop and laptop computers and smartphones, where companies significantly rely on technological devices for online scheduling, routine operations management, and internal and external communications; moreover, they use several applications (e.g., Twitter, Facebook, and LinkedIn) for interfacing with prospective job applicants, consumers, and other stakeholders to answer questions, message, and undertake media advertising (Castille & Sheets, 2012; Miller-Merrell, 2012)

The aforementioned dependency on modern technology has brought challenges for organizations in recent years, with numerous inadvertent organizational consequences.

Moreover, the growth of digitalization leads to clogged bandwidth due to the misallocation of online connectivity, Internet misuse, adverse employee behaviors, and illegal employee acts (e.g., downloading banned or copyrighted information); as such, organizations face several financial problems and risks (Castille & Sheets, 2012; Li & Lin, 2018; Ter Hoeven et al., 2016).

The repercussions of technology dependency produce multitudinous personal problems, namely, higher levels of distraction in the workplace, lower levels of work productivity, improper use of personal information, privacy invasion, technology addiction, and eyesight impairment; furthermore, technology brings another difficulty for employees, who are anxious and irritated when they cannot use their mobile phones, lose their internet connectivity, or simply that their phone is far from their side (Bhattacharya et al., 2019; Duke & Montag, 2017; Li & Lin, 2018; Lupo et al., 2020; Ter Hoeven et al., 2016).

Because of such anxiety and irritation (i.e., nomophobia), individuals cannot concentrate on their tasks or perform well (Alavi et al., 2020; Bian & Leung, 2015).

The workforce is undergoing significant changes because of technological advancements, presenting fresh prospects and obstacles for employers and employees. Businesses must adapt to these developments and equip their employees with the essential skills and tools to thrive in a fast-changing environment. On the other hand, employees must also be ready to acquire new skills and adjust to emerging technologies to stay competitive in the job market (Arlitt et al., 2023; Rotatori et al., 2021).

Accordingly, literature to date encourages research to investigate the effects of the changing workforce due to new technologies (e.g., smartphones, the internet, and social media) (Colbert et al., 2016).

There is little doubt nomophobia creates not only various problems for the individual in their personal life but also challenges in their workplace. Thus, there is a need to investigate the impacts of nomophobia on employment (Wang & Suh, 2018). Furthermore, several studies concentrate on the antecedents and repercussions of this new phobia, even though there is a pivotal need to assess its inhibitors (Rodríguez-García et al., 2020). Nomophobia is a modern phobia caused when individuals are unable to use their smartphones or other technological devices (e.g., smartwatch, laptop, tablet), and is classified in four (4) dimensions as introduced in the following statements according to Yildirim & Correia (2015):

<u>Not being able to communicate</u>: "The feelings of losing instant communication with people and not being able to use the services that allow for instant communication."

<u>Losing connectedness</u>: "The feelings of losing the ubiquitous connectivity smartphones provide, and being disconnected from one's online identity, especially on social media."

<u>Not being able to access information</u>: "The discomfort of losing pervasive access to information through smartphones, being unable to retrieve information through smartphones and search for information on smartphones."

<u>Giving up convenience</u>: "The feelings of giving up the convenience smartphones provide and reflect the desire to utilize the convenience of having a smartphone." (Yildirim & Correia, 2015).

Drawing upon the outlined narrative, it is evident that nomophobia, catalyzed by the extensive integration of digital devices in organizational frameworks, is escalating as a notable concern with discernible ramifications in contemporary workplaces. The anxiety and distraction induced by nomophobia not only deter individual efficacy but may also incite a broader organizational discord, embodying issues like internet misuse, improper conduct, and fiscal strains. The intertwining of nomophobia, social media addiction, and other psychological facets further complicates the scenario, advocating for a deeper scrutiny within modern literature (Tuco et al., 2023). The transitional workforce, emblematic of a significant digital shift, necessitates a balanced approach that capitalizes on technological advantages while concurrently addressing and mitigating the negative implications of nomophobia. The prevailing literature, showcasing a blend of epidemiological, diagnostic, and psychological insights into nomophobia, emphasizes an urgent call for more exhaustive research to unravel the antecedents, consequences, and potential deterrents of this contemporary phobia, aspiring to cultivate a conducive work ambiance in the digital epoch.

## Ethical Environment

Ethical environment refers to the organizational atmosphere regarding the ethical content of activities or aspects of work that relate to ethical behavior. It is about the sense of whether we are doing things right or the feeling that drives us toward a specific type of behavior. In organizations, the ethical climate serves as an important source of information for employees to consider what actions are "right" or ethical in a work context (Victor & Cullen, 1988). Therefore, the perceived ethical climate helps individuals in determining issues related to ethics and deciding which criterion to use for understanding, evaluating, and addressing these ethical matters (Barnett & Vaicys, 2000).

In other words, the ethical climate offers a specific trait of an organization and can potentially change or enhance working conditions. It also represents the organization's procedures, performance, and policies in an ethical context. The cultural and social environment, organizational psychological climate, organizational history, ethical codes and standards, ethics training programs, management practices, and communication methods are all factors contributing to shaping the ethical climate in an organization (Philippe, 2007). Organizational ethical climate is one of the dimensions of organizational climate (Victor & Cullen, 1988).

Organizational climate is regarded as the personality of an organization, and the ethical climate, as a part of this personality, somewhat reflects the ethics of that organization. Nowadays, organizations are striving for survival and greater profitability, making it crucial consequences (Gillespie et al., 2008). The increased emphasis on understanding employees and their behaviors within organizations has led to specific attention to topics such as examining employees' perceptions of organizational climate (Riggio, 2007; Kuenzi & Schminke, 2009).

On the other hand, various unethical behaviors emerging within organizations and their harmful consequences have drawn the attention of many managers and organizational authorities to ethics in work environments (Simms & Keon, 1997). Although ethical decision-making models do not universally agree on all characteristics, they all assume that ethical decision-making by individuals in organizations cannot be perceived without considering the context in which the decision-making process occurs.

Therefore, these models encompass not only the influence of individuals on ethical decision-making but also organizational factors such as reward systems, norms, work procedures, and organizational climate (Barnett & Vaicys, 2000). Given these factors, it is evident that one of the necessary strategies for investigating the causes and addressing unethical behaviors of employees within organizations is to examine the organizational climate related to ethical issues.

Organizational climate is defined as "common perceptions of formal and informal organizational policies, practices, and procedures" (Theurer et al., 2018). Many believe that there are numerous organizational climates, such as service climate, safety climate, innovative climate, and so on. Ethical climate is "one of these work climates within organizations" (Peterson, 2002). Based on these findings, organizations are encouraged to establish and maintain organizational climates that incentivize ethical behavior. Therefore, organizational experts and researchers need to understand more about the ethical climate and its relationship with organizational, individual, and behavioral variables.

Understanding the characteristics of an ethical climate can assist organizational experts in designing and implementing programs to enhance awareness of ethical issues and improve the ethical behavior of employees and management (Kinicki, 2003). Several significant studies have shed light on the influential factors shaping ethical actions and behaviors within organizations. One notable research effort involved surveying 1500 executive managers from various companies who were readers of the HBR magazine. These respondents were asked to rank several factors affecting ethical conduct within organizations. The results highlighted some key elements: top management behavior, ethical practices within the industry, colleagues' and peers' conduct within the organization, formal organizational policies, and personal financial needs, as outlined in the study conducted by Barnett and Vaicys in 2000.

In another extensive survey encompassing the opinions of over 1400 managers (Ivcevic et al., 2020), the impact of six pivotal factors on unethical behavior was explored. Although there were slight variations in the rankings across these studies, certain consistent findings emerged. Notably, top management behavior stood out as the most influential factor in all three studies, underscoring the profound impact of leadership on ethicality within organizations. Additionally, colleagues' behavior was highly ranked in two of these studies, indicating the significant influence of peer behaviors and expectations on individual conduct.

Ethical practices within the industry or profession also consistently held substantial importance, emphasizing the contextual factors shaping ethical decision-making.

Surprisingly, personal financial needs were consistently ranked lower in all three studies (Shayari, 2005). Despite this, the significance of personal financial considerations should not be entirely disregarded, suggesting a more nuanced understanding of the interplay between personal and organizational ethics. Of particular interest from an organizational standpoint is the influence of peer and superior behavior on individuals, indicating a strong connection between organizational climate and managerial ethics (Bass, 1985; Brown & Trevino, 2006). The concept of organizational climate, often discussed in these studies, plays a crucial role in shaping ethicality within the managerial realm (Jones & George, 1998).

However, it is noteworthy that this factor ranked lower in two studies (Vermeer et al., 2019; Trevino et al., 2014), suggesting that while it acts as a background influence, it may not directly dictate organizational ethics. Moreover, while personal needs were ranked lower, their presence in the ethical decision-making landscape implies the existence of intervening factors where managerial discretion comes into play (Detert et al., 2007; Trevino et al., 2018).

These findings collectively underscore the multifaceted nature of organizational ethics, highlighting the intricate balance between leadership behavior (Treviño et al., 2014), peer influence (Brown et al., 2020), industry standards (Campbell et al., 2021), formal policies (Greve & Palmer, 2022), and personal considerations (Drescher et al., 2023). Managers, as pivotal decision-makers within organizations, navigate this complex terrain, exercising their discretion within the framework defined by these influential factors (Treviño et al., 2020). Recognizing and understanding these dynamics is essential for both managers and organizations, fostering a comprehensive approach to ethical decision-making that acknowledges the interplay of various factors shaping organizational ethics (Treviño et al., 2018).

Based on the outlined narrative, it is apparent that the ethical climate within organizations plays a pivotal role in guiding both individual and collective behaviors toward ethical conduct (Victor & Cullen, 1988; Barnett & Vaicys, 2000; Umphress et al., 2010; González-Torres et al., 2023; Simha & Cullen, 2012). An ethical climate fosters a shared understanding among employees regarding what constitutes ethical behavior, thereby assisting them in navigating complex ethical dilemmas (Deng et al., 2023; Parboteeah et al., 2018). For instance, a study by Deng et al. (2023) elucidated how organizational ethical self-interest climate could potentially engender unethical behaviors within the organizational setting, shedding light on the nuanced interplay between organizational climate and ethical conduct (Parboteeah et al., 2018). Moreover, a meta-analytical review by Parboteeah et al. (2018) across national contexts has enriched the ethical climate theory by relating it to various forms of organizational climates, such as safety and innovation climates (González-Torres et al., 2023).

Furthermore, the exploration of ethical climate in healthcare settings by Essex et al. (2023) exemplifies how different sectors may necessitate distinct approaches to fostering an ethical climate (Swalhi et al., 2023). In the grander scheme, understanding and cultivating a positive ethical climate is not only integral for promoting ethical behavior but also for enhancing overall organizational performance (Essex et al., 2023; Swalhi et al., 2023). The cumulative evidence accentuates the importance of continuous research and discourse in this domain to ensure organizations are equipped with the knowledge and tools necessary to nurture a conducive ethical climate, ultimately contributing to a more ethically sound and socially responsible business landscape (Swalhi et al., 2023).

## Organizational Commitment

### Definitions of Organizational Commitment

Commitment can be seen as a form of obligation that restricts freedom of action (Oxford Dictionary, 1969). Organizational commitment is a psychological state that provides the inclination, need, and obligation to continue employment within an organization. It encompasses normative, affective, and continuous dimensions, understanding which plays a crucial role in promoting commitment (Allen & Meyer, 1990, p.3). According to Steers, commitment can arise from organizational features such as employees' decision-making freedom and their fundamental sense of security (Steers, 1983). There are numerous reasons why an organization must increase the level of organizational commitment among its members (Steers & Porter, 1992).

Organizational commitment, as a dependent variable, represents a force that binds an individual to remain within an organization and work towards its goals with a sense of belonging (Esmaili, 2001). Organizational commitment is associated with a set of generative behaviors (Meyer et al., 1993). Individuals with high organizational commitment remain within the organization, embrace its objectives, and demonstrate substantial effort, dedication, and even sacrifice to achieve these goals (Mitchell, 1977). The three-component model of organizational commitment has led to extensive empirical research (Allen & Meyer, 1990). According to this model, organizational commitment consists of three components: affective commitment, normative commitment, and continuance commitment (Allen & Meyer, 1990). Industrial and organizational psychologists, as well as human resource managers, have focused on job attitudes due to various reasons (Mitchell, 1977).

Among the major reasons is the need to understand variables and factors that influence work-related behaviors, such as job performance, turnover, attendance, tardiness, and more (Mitchell, 1977). Based on theories that have formed the basis of numerous studies, meaningful relationships exist between job attitudes and work-related behavioral variables (Judge & Kammeyer-Mueller, 2012). For example, studies by Vroom and Yetton (1985) have shown a meaningful relationship between job satisfaction and job performance. These research findings indicating significant relationships between job attitudes and work behaviors have led industrial and organizational psychologists to explore variables that influence job attitudes (Richards, 1985). They add that by obtaining these findings, theories and models concerning job attitudes can be established or further developed (Cohen & Lowenberg, 1990). Moreover, these findings can be used to recommend to human resource managers ways to foster desirable job attitudes, consequently improving work behaviors and reducing turnover intentions (Mathieu & Zajac, 1990). Some of the advanced and influential variables on job attitudes include personal characteristics such as age, gender, marital status, education, employment history, skill level, work ethics, number of children, and job-related characteristics such as skill variety, task identity, task significance, autonomy, and job feedback. Organizational characteristics like organizational size, organizational centrality (organizational focus), role ambiguity, role conflict, role overload, and many other variables are also among the influential variables (Mathieu & Zajac, 1990).

However, it must be acknowledged that managers are not particularly interested in knowing all these attitudes. In fact, managers are more interested in attitudes related to work and the organization (Spector, 1997). According to research conducted in this area, three major attitudes have garnered the most attention and investigation by researchers (Spector, 1997). These attitudes are: 1. Job Satisfaction; 2. Job Involvement; and 3. Organizational Commitment.

The examination of organizational commitment in the provided text illuminates its multidimensional nature (Meyer & Allen, 1990) and pivotal role in influencing job-related behaviors, attitudes, and, consequently, the broader organizational performance (Mathieu & Zajac, 1990). Recent research further enriches understanding of this concept. For instance, a study by Huynh, Bui, and Nguyen (2023) explored how job satisfaction correlates with organizational commitment among educational managers in Vietnam, underlining the importance of job satisfaction in fostering managerial commitment. The pandemic has also reshaped the concept of organizational commitment, illustrating its ever-evolving nature in response to external crises, as highlighted by Chauhan, Howe, and Nachmias (2023). Furthermore, the interplay between work engagement and organizational commitment in enhancing job performance has been underscored, signifying the relevance of employee engagement in nurturing organizational allegiance (Nabhan & Munajat, 2023).

Additionally, the quality of work life (QWL) has been emphasized as a significant determinant of organizational commitment, suggesting that improving QWL can lead to heightened organizational allegiance, which, in turn, contributes to organizational success (Sampath & Sutha, 2011).

In synthesizing these insights, it's evident that organizational commitment remains a critical focus in both academic and practical realms, with its implications extending to job satisfaction, job performance, managerial behavior, and broader organizational efficacy (Allen & Meyer, 1990; Mathieu & Zajac, 1990). The multifaceted nature of organizational commitment, encompassing affective, continuous, and normative dimensions (Meyer & Allen, 1990), necessitates a holistic approach in its investigation and management (Sampath & Sutha, 2011). Contemporary challenges such as the COVID-19 pandemic, job security, and quality of work life further underscore the dynamic interplay of individual, job-related, and organizational factors in shaping organizational commitment (Chauhan et al., 2023). These recent findings reinforce the necessity for organizations and managers to foster a conducive environment that promotes organizational commitment (Nabhan & Munajat, 2023), which, in turn, augments organizational performance and success (Mathieu & Zajac, 1990).

### Attitudes in Social Psychology Literature

In the literature of social psychology, the concept of attitudes holds special significance. Allport (1935), in a review of studies related to attitudes, stated that the concept of attitudes might be one of the most distinct and essential concepts in contemporary social psychology. It is believed that personal attitudes within the organizational context are related to personality, motivation, and other individual processes (Ajzen & Fishbein, 1980). Attitudes refer to the tendency or readiness to respond in a favorable or unfavorable manner (Eagly & Chaiken, 1993). They are emotionally learned evaluations concerning individuals, objects, and concepts present in our surrounding world (Eagly & Chaiken, 1993). Attitudes are intricately woven into the fabric and structure of our psychological makeup. They are related to our core values and reflect our beliefs about the subjects they pertain to (Eagly & Chaiken, 1993). If meticulously assessed and considered alongside other factors, such as social norms, they can serve as powerful predictors of behavior and form the foundation of our knowledge for interacting with others and the world around us (Mitchell, 1977).

There are various attitudes concerning work activities, with some of the most important ones being job satisfaction, organizational commitment, and job involvement. Most research in this area has focused primarily on job satisfaction, followed by organizational commitment (Saal & Knight, 1995).

### Who Possesses Higher Organizational Commitment?

Employees with a strong organizational commitment generally have better and longer service records compared to less committed ones (Mowday et al., 1982). Individuals who stay with an organization for an extended period usually exhibit a robust organizational commitment (Steers, 1977). In general, those with more experience likely possess higher trust and competence in their jobs, displaying positive behavior and feelings towards the organization (Meyer & Allen, 1990).

Higher levels of the organization also tend to have higher organizational commitment compared to individuals at lower levels (Rezaeian, 2006). This is because positions of power empower individuals to influence organizational decisions. Those in higher positions have more freedom to focus their efforts on their jobs. Higher job positions come with independence and opportunities for involvement and collaboration in decision-making. The ability of high-level employees to make choices increases their sense of control over the environment and leads to loyalty and dedication to the organization (Rezaeian, 2006).

### New Theories and Models of Organizational Commitment

Organizational commitment is a crucial job-related and organizational attitude that has been of great interest to researchers in the fields of organizational behavior and psychology, especially social psychology, in recent decades. This attitude has undergone changes over the past three decades, with perhaps the most significant change being the shift from a one-dimensional perspective to a multidimensional view of this concept.

Additionally, due to recent developments in the business arena, such as mergers and downsizing, some experts have suggested that the impact of organizational commitment on other crucial variables in management, including job turnover, absenteeism, and performance, has diminished (Meyer, 1997). Therefore, some argue that studying it further might be unnecessary. However, others do not subscribe to this view and believe that organizational commitment has not lost its significance. They assert that it can still be studied, considering that employees' behaviors in organizations can be influenced by their attitudes (Rezaeian, 2006). Hence, awareness of these attitudes remains essential for organizational managers (Rezaeian, 2006).

## The Necessity of Attention to Organizational Commitment

There are many reasons why an organization should increase its members' level of organizational commitment. Firstly, organizational commitment is a distinct concept and differs from dependency and job satisfaction. There are several reasons why an organization must enhance the level of organizational commitment among its members (Steers & Porter, 1992: 290). Firstly, organizational commitment is a distinct concept and differs from job satisfaction dependency. For example, nurses may enjoy the tasks they perform (Greenberg & Baron, 2000, p. 182), but they could be dissatisfied with the hospital they work in, leading them to seek similar jobs in similar environments elsewhere. Conversely, restaurant servers might have a positive experience in their work environment but detest waiting tables or their job in general (Greenberg & Baron, 2000: 182). Secondly, research has shown that organizational commitment has positive relationships with outcomes such as job satisfaction (Bateman & Strasser, 1984), attendance (Mathieu & Zajac, 1990), extra-organizational behavior (Armenakis & Bedeian, 1986), and job performance (Meyer, Allen, & Smith, 1993). It also has a negative relationship with turnover intentions (Mowday, Porter, & Steers, 1982) (Shian-Chang et al., 2003: 313) (Rezaeian, 2006).

## Personal Characteristics Influencing Organizational Commitment

**Age**: Organizational commitment is relatively and positively correlated with age (Meyer & Allen, 1990). Many researchers believe that age is more related to calculative commitment, citing the limited opportunities outside the current job and the costs incurred in the later years (Mowday et al., 1982).

**Gender**: Women tend to have higher commitment to organizations compared to men, although the difference is minor (Rezaeian, 2006). The reason for this is that women often have to overcome more obstacles to join an organization.

**Education**: The relationship between organizational commitment and education is weak and negative (Iaffaldano & Muchinsky, 1984). This relationship is more related to affective commitment and does not have a significant connection with calculative commitment. The negative correlation is due to the higher expectations of educated individuals and their greater job opportunities.

# Factors Influencing Organizational Commitment

Some researchers believe that creating organizational commitment is essential due to its connection with work-related behaviors such as absenteeism, turnover, job satisfaction, engagement, performance, and supervisor-subordinate relationships (Meyer & Allen, 1990).

Marital Status

*Marriage*: This variable has a weak correlation with organizational commitment (Mowday et al., 1982). However, it is suggested that marriage is related to calculative commitment due to financial issues (Rezaeian, 2006).

Organizational Tenure and Position

*Tenure*: Due to an individual's investment in the organization, longer tenure in a position or organization leads to higher commitment, although this relationship is weak (Steers, 1977).

*Position*: Job level has a positive but weak relationship with organizational commitment (Sadeghifar, 2007).

Inferencing Personal Merit

Individuals commit to an organization to the extent that it provides opportunities for their growth and success. Therefore, those with higher personal merit expectations will have a stronger positive relationship with organizational commitment (Meyer & Allen, 1990).

Skills

Individuals with high skills are valuable to an organization, enhancing the organization's reward, thus leading to calculative commitment (Mowday et al., 1982).

Salary and Wages

Salary and wages enhance an individual's self-esteem, thus increasing affective commitment. Moreover, salaries and wages are considered a type of opportunity within the organization, which will be lost upon leaving. Various studies have shown a positive but weak correlation between these two variables (Meyer & Allen, 1990).

**Organizational Commitment and Absenteeism**: The relationship between organizational commitment and intentional absenteeism is theoretically inverse. Meyer and his colleagues (1990) found that the connection between affective commitment and intentional absenteeism is inverse. Employees who are emotionally attached to their organization are less likely to take intentional absences.

**Organizational Commitment and Turnover**: The relationship between organizational commitment and turnover is also inverse. Employees who are more committed are less likely to leave the organization compared to those who are not committed. Li and his colleagues (2006) demonstrated that organizational commitment over a four-year period predicted turnover. In a study conducted by Tamimson, it was shown that affective commitment and continuance commitment are negatively related to the intention to quit (Rezaeian, 2006).

**Organizational Commitment and Job Satisfaction**: In the study by Anderson and Oberholtzer in 1992, there are four assumptions about the relationship between organizational commitment and job satisfaction:

- Job satisfaction leads to organizational commitment.
- Organizational commitment leads to job satisfaction.
- Organizational commitment and job satisfaction have a mutual relationship.
- There is no causal relationship between job satisfaction and organizational commitment.

Research by Williams and Hanner (1979) supported the first assumption, studies by Bitman (1998) supported the second assumption, and studies by Corey and his colleagues (1991) supported the fourth assumption (Golipour, 2001).

### Organizational Commitment and Engagement with Work
Engagement with work somewhat indicates job identity and refers to the extent to which an individual measures their self-worth through their job and job performance. Individuals who are engaged in their work have higher performance and lower absenteeism (Kahn, 1990). Organizational commitment has a direct relationship with engagement with work (Kahn, 1990; Rezaeian, 2006).

### Organizational Commitment and Punctuality
Studies have shown an inverse relationship between organizational commitment and employee lateness (Rezaeian, 2006). In other words, more committed individuals strive to be punctual. The study by Engell and Perry (2005) demonstrated a strong inverse relationship between commitment and employee lateness.

### Organizational Commitment and Job Stress
Some believe that affective commitment acts as a buffer to mitigate the negative effects of job-related stress on employees' well-being (Meyer & Allen, 1990). In other words, employees who are emotionally attached to their organization are better able to cope with job stress. Others argue that committed employees might be more vulnerable to such stressors than less committed ones (Modani, 2005). However, continuance commitment is positively correlated with stress and creates a conflict between work and family (Modani, 2005).

### Organizational Commitment and Job Performance
It is predicted that commitment affects the effort an employee puts into their job, and this effort, in turn, influences job performance (Allen & Meyer, 1990). Research has shown that commitment is positively correlated with individual and group performance indicators (Rezaeian, 2006). However, according to Allen and Meyer (1990), employees' inclination to contribute to organizational effectiveness and assist it is influenced by the nature of their commitment.

Employees who are emotionally attached to the organization are more likely to exert effort and contribute to the organization compared to those who only need to belong to an organization (continuous commitment). Therefore, it is noteworthy that studies demonstrating a positive correlation between commitment and performance have often utilized emotional commitment as the commitment indicator. However, it is also possible that the obligation to remain in an organization leads to an obligation to participate in and assist the organization. In this case, continuous commitment might also have a weak positive correlation with performance. Under normal circumstances, employees whose service in the organization is initially based on necessity (continuous commitment) might perceive no reason to do more than what is necessary to maintain their membership in the organization (Rezaeian, 2006).

## The Relationship between Participation and Organizational Commitment

Research findings indicate that one of the influential factors in shaping and strengthening organizational commitment is individuals' participation in organizational affairs and decision-making processes (Allen & Meyer, 1990; Mowday et al., 1982). If employees have genuine involvement in the planning process and goal-setting of the organization and feel that their involvement in decision-making affects their destiny and fulfills their basic needs, they become committed to the organization (Lawler, 1992; Locke & Schweiger, 1979). They recognize the organization's goals and values in alignment with their own objectives and will spare no effort in achieving them (Kanter, 1983; Ouchi, 1981). Therefore, it is recommended that organizational managers provide the necessary platforms for the comprehensive participation of employees.

By addressing employees' issues and problems and by creating facilities and opportunities for them to utilize organizational resources, managers should take charge of directing employee participation. They should engage the organization's members by acknowledging the efforts needed in this regard. In this way, managers can demonstrate their commitment and support for establishing and consolidating employee participation (Stephens, 2001).

## The Relationship between Performance Evaluation and Rewards with Organizational Commitment

According to Bamberg and Moesch (2001), companies that adhere to a committed strategy consider the following indicators in the areas of performance evaluation and rewards:

*Performance-Based Pay*: A meta-analysis by Ng and Nwachukwu (2017) found that performance-based pay has a positive impact on organizational commitment, especially for high-performing employees. This may be because performance-based pay signals to employees that their contributions are valued by the organization and that they are being rewarded for their hard work. Additionally, performance-based pay can motivate employees to achieve higher levels of performance, which in turn can lead to increased organizational commitment.

*Emphasis on Intrinsic Rewards*: A study by Li et al. (2016) found that intrinsic rewards, such as the opportunity for personal growth and development, are more strongly correlated with organizational commitment than extrinsic rewards, such as money and benefits. This may be because intrinsic rewards provide employees with a sense of satisfaction and accomplishment that is not tied to material possessions. Additionally, intrinsic rewards can help employees feel more connected to their work and to the organization as a whole.

*Internal Equity*: A study by Chen and Huang (2018) found that internal equity, or the perception that employees are being rewarded fairly relative to their peers, has a positive impact on organizational commitment. When employees perceive that they are being rewarded fairly, they are more likely to feel valued and respected by the organization. Additionally, internal equity can help to reduce feelings of injustice and dissatisfaction among employees, which can lead to higher levels of organizational commitment.

*External Equity*: A study by Zhou et al. (2019) found that external equity, or the perception that employees are being rewarded fairly relative to employees in similar positions in other organizations, also has a positive impact on organizational commitment. Employees who believe their remuneration is equitable relative to counterparts in different companies tend to experience greater contentment with their salary and incentives. Furthermore, the sense of fair pay contributes to lower staff attrition, with employees being more inclined to stay in their roles when they perceive fairness in their compensation.

*Long-term Evaluation*: A study by DeLong and Beatty (2021) found that long-term performance evaluations are more predictive of employee success than short-term evaluations and are also associated with higher levels of organizational commitment. This may be because long-term performance evaluations provide employees with a more accurate picture of their overall performance and potential. Additionally, long-term performance evaluations can help employees see how their work contributes to the long-term goals of the organization, which can lead to increased organizational commitment.

*Fairness in Data Collection and Performance Evaluation*: A study by Yang and Huang (2020) found that employees who perceive the data collection and performance evaluation process to be fair are more likely to be committed to the organization. When employees perceive that the performance evaluation process is fair and transparent, they are more likely to trust the organization and to feel that their contributions are being valued accurately. Additionally, fairness in performance evaluation can help to reduce employees' feelings of stress and anxiety, which can lead to higher levels of organizational commitment.

## Strategies for Increasing Organizational Commitment

Organizational commitment, as a crucial concept, needs to be expanded within organizations, and more attention should be given to it. Below are some important strategies for enhancing organizational commitment that organizations can consider:

*Increasing emotional attachment among employees and engaging them more with organizational goals*: Employees who feel emotionally attached to their organization are more likely to be committed to it (Rezaeian, 2006). A 2019 study by Khan et al. found that employee emotional attachment to the organization is positively associated with organizational commitment, especially for employees with high levels of job satisfaction.

*Improving social communication networks at work*: Strong social ties between employees can lead to increased organizational commitment (Podsakoff et al., 1996). A 2018 study by Kim and Lee found that social media can play a role in improving social communication networks at work, leading to increased employee engagement and organizational commitment.

*Involving employees in decision-making processes*: When employees feel that they have a voice in the decision-making process, they are more likely to be committed to the organization (Deci & Ryan, 2000). A 2017 study by Jiang and Li found that employee participation in decision-making is positively associated with organizational commitment, especially for employees with high levels of job satisfaction.

*Evaluating employee performance to provide constructive feedback*: Regular performance evaluations help employees to understand their strengths and weaknesses and to set goals for improvement. Constructive feedback from managers can help employees feel valued and supported, which can lead to increased organizational commitment (Locke & Latham, 2002).

*Clarifying organizational goals and missions*: When employees understand the organization's goals and missions, they are more likely to feel invested in achieving them. A 2016 study by Acar found that organizational clarity, which is the degree to which employees understand the organization's goals, vision, and values, is positively associated with organizational commitment.

*Removing workplace obstacles*: Workplace obstacles, such as lack of resources, unclear expectations, and conflict with colleagues, can lead to decreased organizational commitment. A 2018 study by Sharma et al. found that workplace stress is a significant predictor of employee turnover, suggesting that reducing workplace obstacles can help to increase organizational commitment.

*Emphasizing aspects that contribute to the organization's social value*: Employees are more likely to be committed to organizations that they believe make a positive contribution to society (Brown et al., 2005). A 2019 study by Madill and Alvarez found that corporate social responsibility (CSR) programs, which are programs that organizations implement to improve their social and environmental impact, can lead to increased employee engagement and organizational commitment.

*Establishing appropriate reward and punishment systems*: Reward and punishment systems can be used to motivate employees and reinforce desired behaviors. A 2020 study by Zhao et al. found that performance-based pay is positively associated with organizational commitment, suggesting that reward systems that recognize and reward employee performance can help to increase organizational commitment.

*Eliminating discrimination and inappropriate relationships among individuals in the workplace*: Discrimination and inappropriate relationships can create a hostile work environment and lead to decreased organizational commitment. A 2017 study by Lim et al. found that workplace bullying is negatively associated with organizational commitment, suggesting that organizations should implement policies and procedures to prevent and address discrimination and inappropriate behavior in the workplace.

*Allowing relative independence in performing tasks*: Employees who have some autonomy in their work are more likely to be committed to the organization (Hackman & Oldham, 1976). A 2019 study by Park and Joo found that employee autonomy is positively associated with organizational commitment, suggesting that organizations should give employees some control over how they do their jobs.

*Creating a conducive environment for employee creativity and innovation*: Creativity and innovation can lead to improved organizational performance and increased employee engagement. A 2020 study by Zhou and Wang found that a supportive organizational culture, which is a culture that encourages risk-taking, experimentation, and collaboration, is positively associated with employee creativity and innovation.

*Delegating higher levels of responsibility to individuals for job execution*: When employees are given more responsibility, they are more likely to feel challenged and engaged in their work. A 2018 study by Sharma et al. found that job empowerment, which is the degree to which employees have autonomy, control, and responsibility in their work, is positively associated with organizational commitment.

*Ensuring upper management's awareness of employees' level of organizational commitment*: It is important for upper management to be aware of the level of organizational commitment among employees so that they can take steps to improve it. A 2019 study by Khan et al. found that employee-perceived organizational support, which is the degree to which employees believe that the organization cares about them and values their contributions, is positively associated with organizational commitment.

*Promoting internal advancement and reducing factors that lead to a decrease in employees' organizational commitment*: Organizations should promote internal advancement opportunities and reduce factors that lead to decreased organizational commitment, such as burnout, work-life balance problems, and job dissatisfaction. A 2017 study by Jiang and Li found that employee career development opportunities, which are opportunities for employees to grow and develop their skills and abilities, are positively associated with organizational commitment.

## Approaches to Organizational Commitment

Organizational commitment, with the characteristics mentioned, can impact organizational strategies such as competitive advantage and financial success (Brown et al., 2005; Meyer et al., 2002). While being influenced by different factors, understanding them can illuminate the way in the human resources strategic planning of the organization (Jiang & Li, 2017; Khan et al., 2019).

In recent years, research on organizational commitment has focused on both as an independent and a dependent variable. As an independent variable, organizational commitment has been shown to influence employee turnover, job performance, and organizational citizenship behaviors (Podsakoff et al., 1996; Meyer et al., 2002). As a dependent variable, organizational commitment has been shown to be influenced by a variety of factors, including job satisfaction, perceived organizational support, and employee participation in decision-making (Jiang & Li, 2017; Khan et al., 2019). However, most studies have been conducted to understand the influencing factors on organizational commitment. Briefly, research in this area includes:

*Steers and Porter* (1983) found that the initial level of potential employee attachment to the organization, viewed as the primary individual factor, and the ability to participate in job-related decision-making, considered an organizational factor, are influential in shaping organizational commitment (Steers & Porter, 1983). Eventually, they highlighted the ability to access job alternatives as an extra-organizational factor (Steers & Porter, 1991).

*Bateman and Strasser* (1984) measured variables such as leader's rewarding behavior, leader's punishing behavior, job characteristics, need for achievement, external job alternatives, job stress, job satisfaction, age, education, work experience, and career path history in relation to organizational commitment (Bateman & Strasser, 1984).

*Allen and Meyer* (1990), in their research to assess the influential factors in shaping organizational commitment, utilized questions about job challenges, role clarity, goal clarity, goal achievement difficulty, responsibility, coworker cohesion, organization compliance, fairness, job importance for the individual, feedback, and participation (Farhangi, 2005).

## Organizational Commitment Outcomes

Considering the significant impact of organizational commitment on individuals' job behaviors, researchers have focused on its effects. A substantial portion of the literature on this subject emphasizes the positive relationship between organizational commitment and job-related outcomes such as performance (Podsakoff et al., 1996; Meyer et al., 2017), prosocial behaviors (Cropanzano & Mitchell, 2005; Grant, 2017), and organizational citizenship behaviors (OCBs) (Organ et al., 2016; Park & Lee, 2018) (see Table 1). Conversely, a negative relationship has been observed between organizational commitment and behaviors like turnover and turnover intention (Meyer et al., 2002; Zhou et al., 2018), absenteeism and tardiness (Huseman et al., 1990; Maertz & Campion, 2004), and counterproductive work behaviors (CWBs) (Spector et al., 2015; Wang et al., 2018)

# Types of Organizational Commitment

## Affective Commitment:

This aspect of organizational commitment is defined as an emotional attachment to an organization, manifested through the acceptance of organizational values and a willingness to remain in the organization (Meyer & Morin, 2017).

**Preconditions for Affective Commitment**

*Personal Traits*: The importance of individual traits arises because many of these traits play an increasing or decreasing role in commitment. These traits include age, work experience, level of education and expertise, gender, race, marital status, and other personal factors. Age and work experience have a direct relationship with commitment, while level of education has an inverse relationship with organizational commitment (Meyer & Morin, 2017).

*Role-Related Characteristics*: Research indicates that an enriched and challenging job enhances commitment (Podsakoff et al., 1996). Additionally, findings show that commitment has an inverse relationship with role conflict and role ambiguity (Meyer & Morin, 2017).

*Structural Features*: Studies about the organization's size, span of control, centralization, and job complexity show no significant relationship between any of these variables and organizational commitment (Steers, 1977). However, another study by Steers and his colleagues concluded that formality and job complexity were related to commitment (Meyer & Morin, 2017).

*Work Experiences*: Occurring during an individual's work life, they are considered significant forces in the socialization or influence process of employees, affecting their psychological dependence on the organization. Employees' perceptions of positive coworker attitudes have an effective role in individual commitment (Astron, 1998).

## Normative Commitment

Normative commitment is defined as a perceived obligation to support the organization and its activities, expressing a sense of duty and responsibility to remain with the organization (Meyer & Allen, 1991). Individuals believe that continuing their activities and supporting the organization are moral obligations upon them.

**Preconditions for Normative Commitment**

*Familial, Cultural, and Organizational Influence*: Vulnerability to cultural, familial, and organizational influences plays a significant role (Koku, 2023).

*Organizational Investments*: The investments made by the organization in individuals.

*Reciprocal Services*: Mutual compensations in the relationship (Hosseini & Mahdizadeh, 2006).

## Continuous Commitment

Continuous commitment arises from the perception of increasing costs incurred by leaving an organization. Sunk costs refer to expenses of an activity or project that cannot be recovered. Therefore, if someone has continuous commitment, they will be sensitive to the increase in such costs (Meyer & Allen, 1991).

**Preconditions for Continuous Commitment**

*Volume and Scale of Investments and Resources Deployed in the Organization*: The size and scale of investments and resources allocated in the organization (Spekman, 2023).

*Lack of Job Opportunities Outside the Organization*: The absence of job opportunities outside the organization (Rousseau, 1995).

## Methodology

In this study, we have meticulously crafted a robust conceptual framework (refer to Figure 1) by delving into extensive literature assessments and theories. This framework serves as the backbone for our research, guiding our exploration and providing the foundation for the hypotheses we propose for rigorous testing and analysis. To ensure the legitimacy and ethical integrity of our research, a formal letter detailing the study's objectives was meticulously drafted and sent to various esteemed Jordanian business and entrepreneurship schools for approval. A paramount aspect of our study was the assurance of strict anonymity and confidentiality, with an absolute commitment to non-disclosure of any individual data.

In research, the issue of nonresponse bias looms large, arising from disparities between survey participants and those who abstain. This mismatch can skew results, rendering them inaccurate due to differing characteristics of respondents and non-respondents. To counter this, researchers deploy a range of tactics, such as follow-up reminders and incentives, alongside careful sampling techniques. These measures aim to boost response rates and minimize disparities, as noted by Armstrong and Overton (1977). We were gratified to receive approval from five reputable institutions. Subsequently, our dedicated team of researchers approached a substantial sample size of 700 faculty members across these institutions. Throughout our interactions, we placed significant emphasis on the maintenance of anonymity and confidentiality, garnering the trust and cooperation of 650 teachers who willingly participated in our study. This extensive surveying process took place over a period of nearly two months, commencing on March 1st and concluding on April 20th, 2022. Remarkably, our efforts resulted in an impressive 85% response rate, with 600 meticulously completed questionnaires gathered for thorough analysis.

Utilizing Confirmatory Factor Analysis (CFA), we meticulously examined the dimensions of our study scales. Employing descriptive statistics and a correlation matrix, we ensured the reliability and validity of our data. Finally, Structural Equation Modelling (SEM) was harnessed to scrutinize our hypotheses and evaluate the quality of our model. Through these rigorous methods, we fortified the integrity of our study, ensuring robust and reliable findings.

Figure 1: Conceptual framework

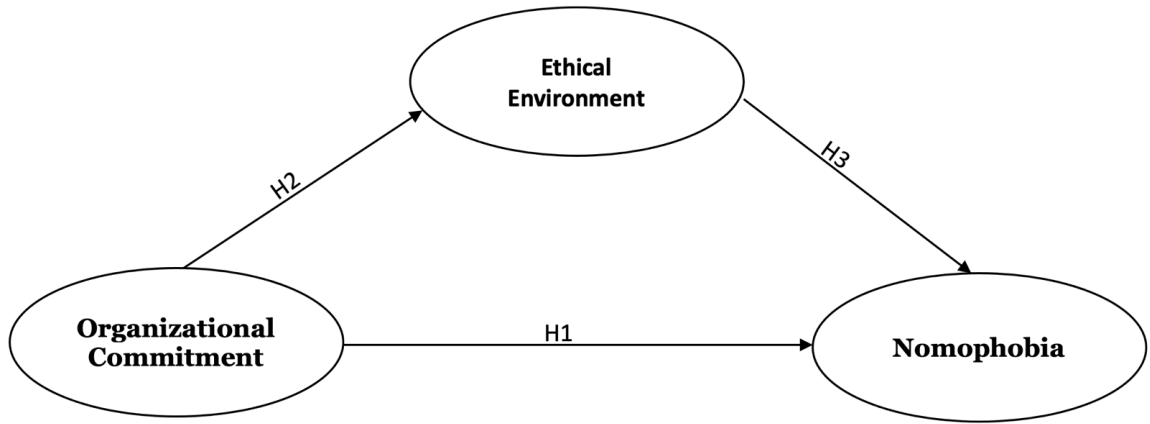

## Measures of the study

In this research, a series of thoughtfully constructed questionnaires was deployed to extract invaluable insights, ensuring a comprehensive exploration of the research variables. To delve into the ethical dimensions, a 6-item questionnaire, rooted in Deshpande's Ethical environment (1996), was expertly administered. This tool adeptly probed the ethical landscape, illuminating the intricate interplay of values and principles within the study's context. Furthermore, the study delved deep into the realm of organizational commitment, employing the comprehensive questionnaire devised by Allen and Mayer (1990). This instrument, comprising eight questions each for affective commitment, normative commitment, and continuous commitment, meticulously assessed participants' dedication to their organizations.

To gain a holistic understanding, participants' nomophobia—the fear of disconnection from smartphones and the internet—was measured through a robust 20-item questionnaire developed by Yildirim and Correia (2015) (NMP-Q). This nuanced instrument thoroughly explored various facets of nomophobia. All responses, pivotal to the study's insights, were painstakingly collected through a refined data-gathering process. Participants articulated their perspectives using a precise "5-point Likert scale ranging from 1= strongly disagree to 5 = strongly agree. "This meticulous approach guaranteed the nuanced and accurate representation of participants' viewpoints, enabling a comprehensive analysis of the gathered data. By adeptly employing these diverse questionnaires and scales, this study painted a detailed and multifaceted portrait of participants' ethical, nomophobia, and job-related commitment. This enriched the depth and validity of the research findings, presenting a nuanced understanding of the intricate dynamics at play.

# Data Analysis Overview

## Measurement Model

Our validation process utilized a consecutive strategy employing Confirmatory Factor Analyses (CFA) to assess the structure of our observed variables. To effectively handle the high multicollinearity between constructs, we opted for CFA, which is well-suited for Covariance-based Structural Equation Modeling (CB-SEM), as suggested by Hair et al. (2021). The analysis of our reflective constructs was conducted utilizing the lavaan package in R Studio, as advocated by Rosseel (2012). Additionally, certain aspects of our analysis were complemented by using SPSS, ensuring a comprehensive and rigorous evaluation of our data.

## Reliability and Validity

Reliability: A stringent evaluation of the research instruments revealed unwavering reliability. The internal consistency of the items was verified using both Cronbach's alpha (CA) and composite reliability (CR), both of which exhibited values surpassing the 0.7 benchmark, signifying the robustness of the measurement items (refer to Table 2). This meticulous scrutiny ensures that the items within each construct reliably measure the intended theoretical concepts, providing a solid foundation for subsequent analyses and interpretations. Convergent Validity: The study's commitment to methodological rigor extended to the assessment of convergent validity.

Through a meticulous examination, all factor loadings were found to exceed the 0.70 threshold, affirming the constructs' ability to capture the underlying dimensions they were designed to measure. Furthermore, the Average Variance Extracted (AVE) values, which surpassed 0.50, as evidenced in Table 2, reinforced the convergent validity of the constructs. This robust validation assures that the measurement items are indeed converging to measure the same construct, substantiating the accuracy and consistency of the research findings. These results underscore the reliability and validity of the measurement model, affirming the integrity of the research methodology.

Discriminant Validity: Rigorous analysis was conducted to establish discriminant validity among the latent variables. Variables with factor loadings exceeding 0.50 were meticulously identified and utilized to validate their distinctiveness from other variables (refer to Table 3). This meticulous scrutiny ensures that each variable stands apart from others, confirming their unique contribution to the research model. The meticulous attention to discriminant validity bolsters the robustness of the study, assuring that the constructs under consideration are not only reliable and convergent but also distinct from one another, thereby enhancing the overall quality and credibility of the research outcomes.

Table 2: Reliability

| Constructs | CR | AVE | Cronbach's alpha | Mean | SD |
|---|---|---|---|---|---|
| Organizational commitment | 0.87 | 0.79 | 0.92 | 2.79 | 0.96 |
| Ethical environment | 0.78 | 0.68 | 0.89 | 2.84 | 1.10 |
| Nomophobia | 0.81 | 0.72 | 0.91 | 2.85 | 1.16 |

Table 3: Discriminant Validity.

|  | Organizational commitment | Ethical environment | Nomophobia |
|---|---|---|---|
| Organizational commitment |  |  |  |
| Ethical environment | 0.617 |  |  |
| Nomophobia | 0.314 | 0.340 |  |

## Measurement model analysis

We utilized confirmatory factor analysis (CFA) in the R Studio software, a widely acknowledged method in the fields of social research and Information Systems (Kline, 2015). The validation of our model adhered to established criteria, with CFI, NFI, and NNFI values exceeding 0.9, while SRMR and RMSEA values remained below 0.08, aligning with the standards set by Hair (2019). Our results demonstrated a strong fit for our hypothesized model, with RMSEA at 0.033, NFI at 0.9, CFI at 0.910, SRMR at 0.031, NNFI at 0.921, and TLI at 0.931. These findings affirm the robustness and adequacy of our model.

## Structural equation modeling results

Employing the lavaan package in R Studio, we meticulously tested the structural equation of our model, integrating control variables into our analysis. The indices evaluated for model fit include the Comparative-Fit Index (CFI), Tucker-Lewis Index (TLI), Normed Fit Index (NFI), and Non-Normed Fit Index (NNFI), which yielded satisfactory values of 0.904, 0.893, 0.915, and 0.939, respectively. Furthermore, our assessment considered the Standardized Root Mean Square Residual (SRMR) and the Root Mean Square Error of Approximation (RMSEA), which registered values of 0.031 and 0.032, meeting the established criteria (Hair et al., 2019). The Relative Chi-Square value (chi-square/degrees of freedom) stood at 1.3, aligning with literature-backed standards for good model fit (Kline, 2015). Our model not only met but exceeded these benchmarks, demonstrating an appropriate fit for our data. This validation is visually represented in Figure 2, showcasing the SEM outcomes derived from R Studio software.

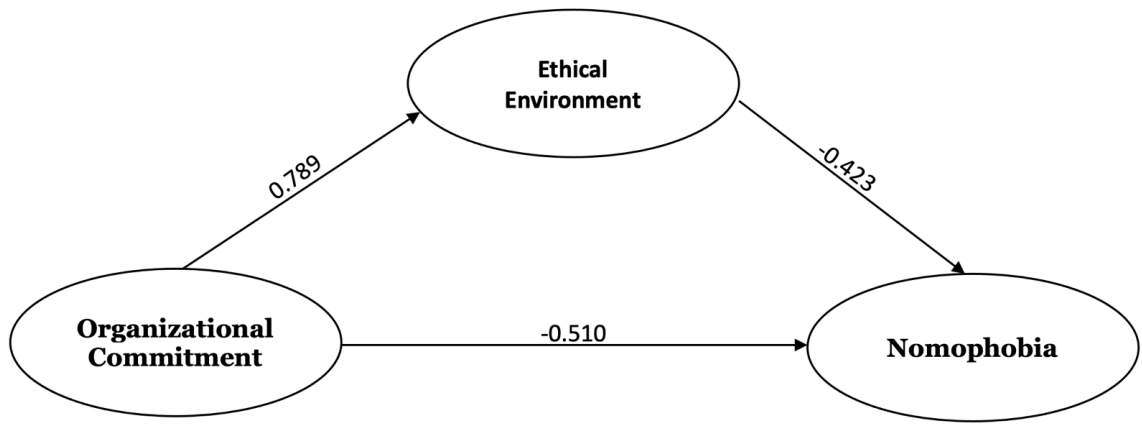

Figure 2: Structural models with standardized estimates.

## Summary of Findings

**Organizational commitment and Nomophobia**: Our analysis, detailed in Table 4, unequivocally supports the first hypothesis, indicating a negative significant influence of organizational commitment on nomophobia (p < 0.000). This finding underscores the substantial positive impact of organizational commitment in reducing employees' nomophobia. Ethical Environment and nomophobia: Furthermore, the second hypothesis, exploring the connection between the ethical environment and nomophobia, was similarly affirmed (p < 0.000). This result underscores the pivotal role played by the ethical climate within an organization in decreasing employees' overall nomophobia. The impact of organizational commitment on the ethical environment was strongly supported (p < 0. 000), indicating a significant enhancement of the ethical climate within the organization due to heightened organizational commitment.

Table 4: Hypothesis Testing.

| Path | Standardized coefficient | P-value | Result |
|---|---|---|---|
| Organizational commitment -> Nomophobia | -0.510 | 0.000 | Supported |
| Ethical Environment -> Nomophobia | -0.423 | 0.000 | Supported |
| Organizational commitment -> Ethical environment | 0.789 | 0.000 | Supported |

**Mediating Role of Ethical Environment**: Extending our investigation to the mediating effects, Table 5 illuminates the fourth hypothesis, confirming that the ethical environment indeed mediates the relationship between organizational commitment and nomophobia (p < 0.000). This nuanced understanding highlights the intricate interplay between organizational commitment, ethical environment, and nomophobia.

Table 5: Mediating Hypothesis Testing

| Path | Mediating | Direct | Indirect | Total | Result |
|---|---|---|---|---|---|
| Organizational commitment -> Nomophobia | Ethical Environment | -0.510 | -0.218 | -0.728 | Supported |

# Discussion

The current paper delves into the intricate relationship between organizational commitment and nomophobia against the backdrop of the ethical environment acting as a mediating factor. The theoretical framework, grounded in Meyer and Allen's three-component model, provides a comprehensive lens for examining how an employee's bond with their organization can mitigate the psychological discomfort associated with nomophobia. This research contributes a critical perspective by highlighting that increased organizational commitment, particularly its affective and normative components, can serve as a protective barrier against the anxiety triggered by technological disconnection (Erdurmazlı et al., 2019). Additionally, this study sheds light on the mediation role of work-family conflict in the relationship between nomophobia and organizational identification(Erdurmazlı, 2022), suggesting that other mediating factors could also significantly impact this relationship.

This study further unravels the role of the ethical environment as a significant mediator, suggesting that a positive ethical climate within an organization can reduce the prevalence and impact of nomophobia on employees. This underlines the dual role of the ethical environment: fostering ethical conduct and simultaneously providing psychological comfort against the stresses associated with constant digital connectivity (Kim & Vandenberghe, 2021). The theoretical development of the study emphasizes the need for a balanced approach to technological integration in workplaces that promotes both the advantages of digital tools and the well-being of employees.

The pervasive integration of technology in the workplace, as the study indicates, is not without its challenges. Nomophobia, in conjunction with social media addiction and related psychological issues, demands a nuanced approach that encompasses both the benefits and the potential drawbacks of our increasing reliance on digital devices (León-Mejía et al., 2021; Kaviani et al., 2020). The research suggests that the modern organization must navigate these complexities by developing strategies that bolster organizational commitment and foster a supportive ethical environment. Moreover, the stress generated from smartphone withdrawal, as demonstrated by Tams et al. (2018), underscores the urgency for organizations to address nomophobia by enhancing organizational commitment and promoting a conducive ethical environment.

# Contribution to the Literature

This study enriches the organizational behavior literature by empirically examining the nexus between organizational commitment and nomophobia, a relationship scarcely addressed in previous research. By leveraging Meyer and Allen's model it extends the theoretical understanding of how commitment dimensions interact with technological stressors. The research further broadens the discourse by introducing the ethical environment as a mediator, thus offering a unique intersection between organizational ethics and technological impact on employees. This integration of multiple theoretical domains presents a more dynamic understanding of the challenges posed by digital dependence in the workplace.

## Contribution to the Practice

From a practical standpoint, this research provides actionable insights for organizational leaders. The demonstrated inverse relationship between organizational commitment and nomophobia reinforces the importance of nurturing a strong organizational bond. Initiatives aimed at enhancing affective and normative commitment are likely to be particularly effective in mitigating the negative psychological effects associated with technological omnipresence. Additionally, the pivotal role of the ethical environment in reducing nomophobia underscores the necessity for ethical leadership and robust ethical policies that can serve as a foundation for employee well-being in the digital era.

## Conclusion

The study's theoretical and empirical examination of nomophobia within organizational settings underscores the complex interplay between organizational commitment, ethical environment, and employee adaptation to technological ubiquity. The findings articulate the protective role of organizational commitment, suggesting that it can diminish the anxiety associated with nomophobia, thereby promoting a more resilient workforce.

These insights compel organizations to consider the human elements within their technological strategies actively. By fostering a culture of strong commitment and ethical clarity, organizations can mitigate the potential psychological discomforts brought on by our digital dependencies. As the workplace continues to evolve with technological advancements, the need for a balanced approach that prioritizes both employee well-being and technological efficacy becomes ever more pressing.

In closing, this research offers a valuable addition to both scholarly discourse and organizational practice. It calls for a renewed focus on the psychological contracts between employers and employees, urging organizations to cultivate environments where technology serves as a complement to a supportive and ethically sound workplace culture.